\documentclass[twocolumn,...]{revtex4}
\usepackage{graphicx}
\usepackage{dcolumn}
\usepackage{bm}
\usepackage{amsmath}
\usepackage{amssymb}
\usepackage{epsfig}\usepackage{epstopdf}
\usepackage{array}
\usepackage{natbib}
\usepackage{ulem}
\usepackage{color} 

\begin{document} 
\title{Anisotropic optical conductivity of the putative Kondo insulator CeRu$_4$Sn$_6$} 

\author{V.\ Guritanu$^1$, P.\ Wissgott$^2$, T.\ Weig$^3$, H.\ Winkler$^2$, J.\
Sichelschmidt$^1$, M.\ Scheffler$^3$, A.\ Prokofiev$^2$, S.\ Kimura$^4$, T.\
Iizuka$^4$, A.\ M.\ Strydom$^5$, M.\ Dressel$^3$, F.\ Steglich$^1$, K.\
Held$^2$, and S.\ Paschen$^2$}
\affiliation{$^1$Max Planck Institute for Chemical Physics of Solids, 01187 Dresden, Germany} 
\affiliation{$^2$Institute of Solid State Physics, Vienna University of Technology, Wiedner Hauptstr. 8-10, 1040 Vienna, Austria}
\affiliation{$^3$1.\ Physikalisches Institut, Universit\"{a}t Stuttgart, 70550 Stuttgart, Germany}
\affiliation{$^4$UVSOR Facility, Institute for Molecular Science, Okazaki 444-8585, Japan}
\affiliation{$^5$Physics Department, University of Johannesburg,
Auckland Park 2006, South Africa}

\begin{abstract}

Kondo insulators and in particular their non-cubic representatives have remained
poorly understood. Here we report on the development of an anisotropic
energy \textcolor{black}{pseudo}gap in the tetragonal compound CeRu$_4$Sn$_6$
employing optical \textcolor{black}{reflectivity} measurements in broad frequency
and temperature ranges, and local density approximation plus dynamical mean
field theory calculations. The calculations provide evidence for a Kondo insulator-like response within the $a-a$ plane and a more metallic response along the $c$ axis and qualitatively reproduce the experimental observations, helping to identify their origin. 

\end{abstract}

\date{\today}

\maketitle

Correlated materials with gapped or pseudo-gapped ground states continue to be
of great interest. The gap in the electronic density of states (DOS) either
opens gradually with decreasing temperature, as the pseudogap of
high-temperature superconductors \cite{Din96.1}, or emerges at a continuous or
first order phase transition \cite{Cru08.1,Sto04.1,Sch10.2}. In heavy fermion
compounds \cite{Col07.1} -- systems in which $f$ and conduction electrons
strongly interact -- a narrow hybridization gap is known to emerge gradually
\cite{Lev11.1,Par12.1,Ass99.1,Ots09.1}. Generically, the Fermi energy is
situated in one of the hybridized bands and a metallic heavy fermion ground
state arises. Only for special cases the Fermi energy lies within the gap and
the ground state is Kondo insulating. Metallic heavy fermion systems have been
intensively investigated over the past decades and are now, at least away from
quantum criticality \cite{Loe07.1}, well understood \cite{Ste84.1} within the
framework of Landau Fermi liquid theory. Hence, a very few parameters, most
notably the effective mass, allow us to describe thermodynamic and transport
properties at the lowest temperatures. In comparison, the physics of  Kondo
insulators has proven to be much less tractable. This is at least in part due to the fact
that the gapped ground state inhibits a characterization via the above
properties. Many experimental efforts have therefore focussed on the
determination of the gap width from temperature dependencies, which has frequently led to conflicting results, in particular for anisotropic Kondo
insulators such as CeNiSn \cite{Tak90.1}. Here the strongly anisotropic
transport and magnetic properties have been interpreted phenomenologically on
the basis of a V-shaped DOS \cite{Kyo90.1} or by invoking a hybridization gap with nodes \cite{Ike96.1,Mor00.1,Yam12.1} or extrinsic effects such as
impurities, off stoichiometry or strain \cite{Tak96.1,Sch92.1}. To advance the
field it appears mandatory to model a number of carefully chosen materials {\it ab initio}, taking all essential ingredients into account.

Here we investigate a new material, CeRu$_4$Sn$_6$, which due to its tetragonal
crystal structure is simpler than the previously studied orthorhombic materials.
In a combined experimental and theoretical effort we provide direct
spectroscopic evidence for the development of an anisotropic
\textcolor{black}{pseudo}gap: While \textcolor{black}{weak metallicity} prevails in
the optical conductivity along the $c$ axis, insulator-like behavior
\textcolor{black}{without a Drude peak} is observed in the $a-a$ plane. We trace
this back to a correlated \textcolor{black}{band structure} which is essentially
gapped, except for the $c$ direction, as shown by local density approximation
(LDA) plus dynamical mean field theory (DMFT) calculations.

CeRu$_4$Sn$_6$ crystallizes in
the tetragonal I$\bar 4$2m structure \cite{Ven90.1, Das92.1}, with $a = 6.8810$ \AA\ and
$c = 9.7520$ \AA. Single crystals were grown from self flux, using the floating
zone melting technique with optical heating \cite{Pas10.1}.  Near-normal incidence reflectivity spectra on the {\it ac}-plane were measured using linearly polarized light, with the electric field
$E \parallel a$ and $E \parallel c$, in a broad energy range from 0.5 meV to 30
eV. In the terahertz (THz) energy range ($0.5-5$~meV) a
coherent-source spectrometer was used \cite{Gor07.1}, for 5~meV-0.68~eV a
Fourier transform spectrometer (Bruker IFS 66 v/S), with a reference gold layer
evaporated {\it{in situ}} on the sample surface, for $0.6 - 1.25$~eV a JASCO
FTIR 610 spectrometer with an Al mirror as reference, and for  $1.2 - 30$~eV
synchrotron radiation at the beamline 7B of UVSOR-II in Japan. Between 5 meV
and 0.68 eV, a magnetic field of 7~T applied to polycrystalline CeRu$_4$Sn$_6$
did not change the reflectivity appreciably.
\begin{figure}[tbh]
\centerline{\includegraphics[width=\columnwidth,clip=true]{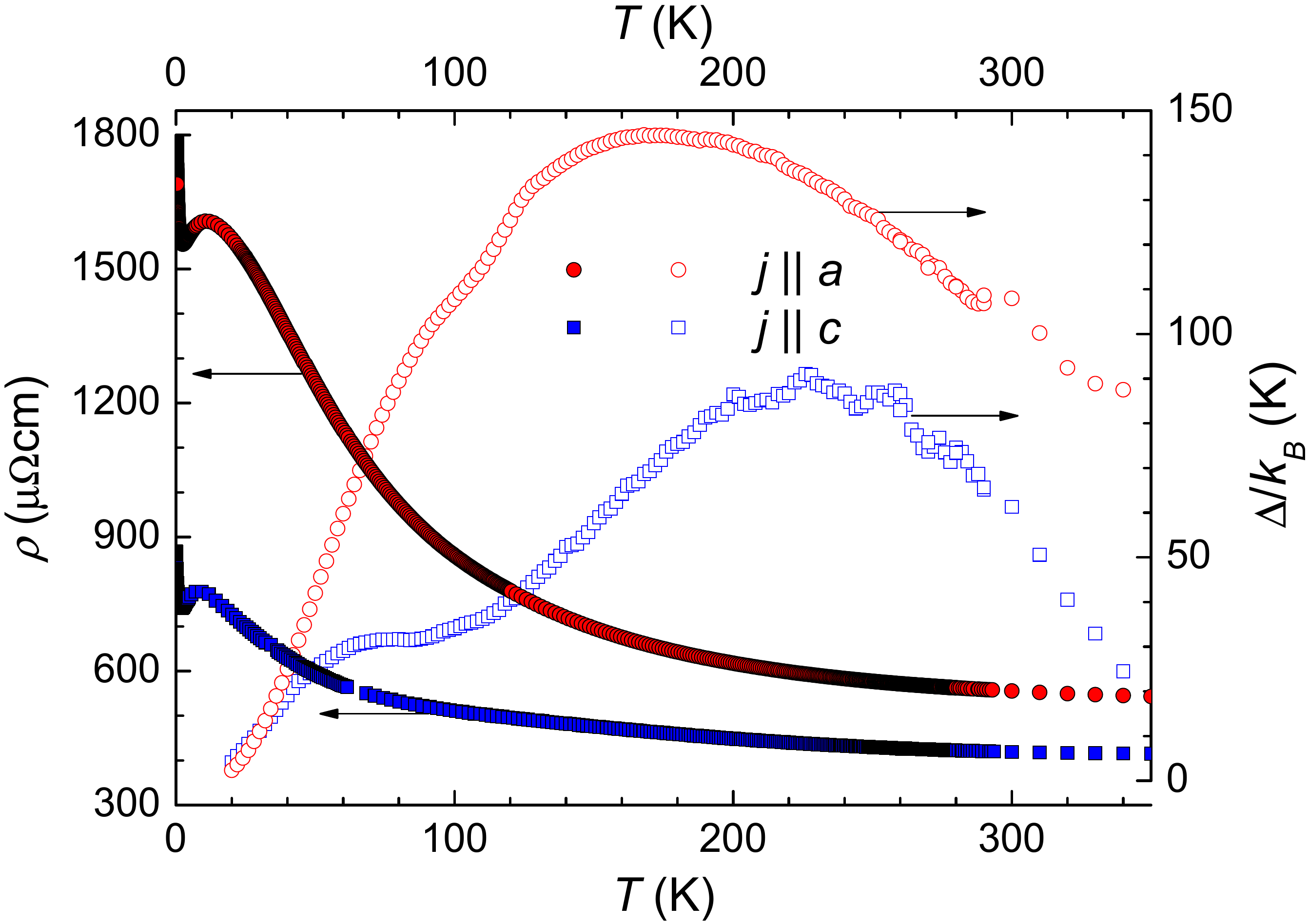}}
\vspace{-0.2cm}

\caption{(Color online) Temperature dependent electrical resistivity, $\rho(T)$
(left), and energy gap in units of temperature, $\Delta/k_B$ (right, see text),
for single crystalline CeRu$_4$Sn$_6$, with the electrical current density $j$
along the $a$ and $c$ axes. } 
\label{rho} 
\end{figure}

\begin{figure*}
{\includegraphics[width=\textwidth]{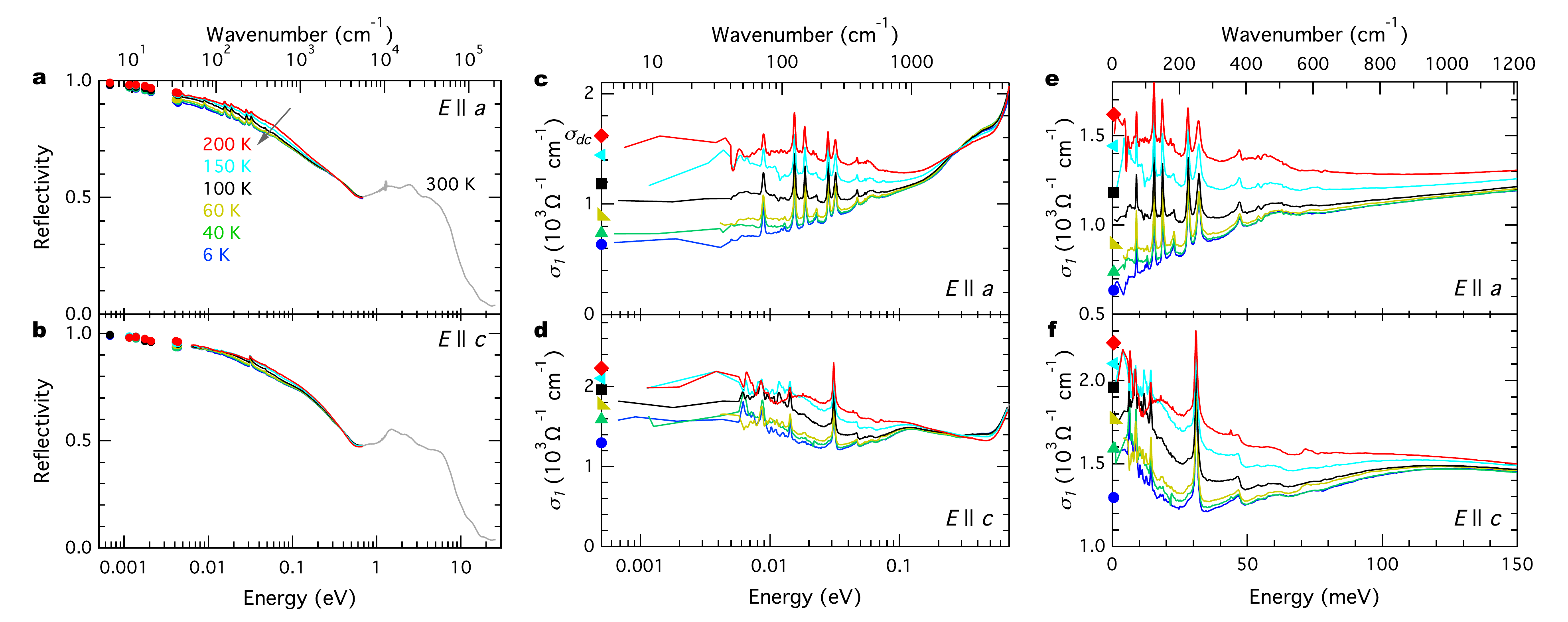}}
\vspace{-0.7cm}

\caption{ (Color online) (a, b) Normal-incidence reflectivity spectra of
single crystalline CeRu$_4$Sn$_6$ at various temperatures for $E \parallel a$
and $E \parallel c$, respectively. (c, d) Real parts of the optical
conductivity, $\sigma_1(\omega)$, for $E \parallel a$ and $E \parallel c$,
respectively. The filled circles on the vertical axis represent the
independently measured dc conductivity data at the corresponding temperatures.
(e, f) Low-energy part of $\sigma_1(\omega)$, showing, at
the lowest temperatures, semiconductor-like behavior and  \textcolor{black}{weak
metallicity} for $E \parallel a$ and $E \parallel c$, respectively. The sharp
features in (c, f) (at 8.9, 15.5, 18.6, 28, 31.6 meV along the {\it a}-axis,
and at 8.7, 14.4, 30.9 meV along the {\it c}-axis) are phonon modes. They are
temperature independent within our resolution ($2\,\rm{cm}^{-1} \equiv
0.25\,\rm{meV}$).}
\label{fig2} 
\end{figure*}

The theoretical optical conductivity was derived from DMFT combined with LDA
within density functional theory \cite{Kot06.1,Hel07.1}. We performed a
full-potential Wien2k \cite{Bla90.1} LDA calculation with spin-orbit coupling,
and projected the Bloch waves onto maximally localized Wannier orbitals of
Ce-$4f$, Ru-$4d$, and Sn-$5p$ character \cite{Kun10.1,Mos08.1}. For this basis
and a typical Coulomb repulsion of $U=5.5$~eV for Ce \cite{Mcm03.1}, we included
electronic correlations by DMFT, using quantum Monte Carlo simulations
\cite{Hir86.1} for the $J=5/2$ subset of the Ce-$4f$ orbitals around the Fermi
energy. The optical conductivity was calculated as described in
Ref.~\onlinecite{Wis12.1}, with an additional imaginary part of the self energy
of 50 meV to account for impurity scattering and a minor readjustment of the
chemicial potential by 50 meV due to different $k$-meshes employed.

As mentioned above, attempts to characterize Kondo insulators by a single,
temperature-independent energy gap have generally failed. This is also the case for CeRu$_4$Sn$_6$, as illustrated by temperature-dependent electrical
resistivity, $\rho(T)$, data along the two principal axes $a$ and $c$
(Fig.\,\ref{rho}, left axis). As for other non-cubic systems (e.g., CeNiSn
\cite{Nak96.1}, U$_{2}$Ru$_{2}$Sn \cite{Tra03.1}, CeFe$_2$Al$_{10}$
\cite{Kim11.2}), a pronounced anisotropy is observed. The
temperature-dependent energy gaps $\Delta(T)$ for the two directions (Fig.\,\ref{rho}, right axis) were obtained by fitting the $\rho(T)$ data in
20~K ranges with an Arrhenius law $\rho=\rho_0\exp(\Delta/(2 k_B
T))$. Our results confirm that the definition of a unique energy gap scale
characterizing the material seems arbitrary. Modeling temperature-dependent data
with fine-structured (V-shaped or more complex) free electron bands yields
improved fits \cite{Bru10.1}, but provides little new insight. This clearly
calls for a rethink of the problem and a novel approach, which is what we present here.

Optical \textcolor{black}{spectroscopy is} a powerful tool to characterize strongly
correlated materials since the low energy scales in these systems can be ideally
probed by optical excitations in the far infrared and THz
frequency range. The optical reflectivity spectra, $R(\omega)$, of CeRu$_4$Sn$_6$ develop
sizable temperature dependence at low energies (Fig.\,\ref{fig2} a, b). This
effect is larger for the {\it a} axis, in agreement with the steeper increase of
the $a$-axis $\rho(T)$ with decreasing temperature (Fig.\,\ref{rho}).

The frequency-dependent real part of the optical conductivity,
$\sigma_1(\omega)$, was derived from the reflectivity data using a
Kramers-Kronig fitting procedure \cite{Kuz05.1,THz}. $\sigma_1(\omega)$ shows
pronounced $a$-$c$ anisotropy (Fig.\,\ref{fig2} c-f), in particular at low
temperatures. At 6~K, $\sigma_1(\omega)$ decreases continuously with decreasing
frequency for the $a$ axis. For the $c$ axis a much more complex frequency
dependence is observed: $\sigma_1(\omega)$ decreases down to 0.5 eV, passes over
a local maximum at 120 meV and a local minimum at 30 meV, and increases with
further decreasing frequency. At lowest frequencies, the optical conductivity is
almost frequency independent for both polarizations and satisfying agreement
with the dc values determined from $\rho(T)$ is found. 

While the continuous depletion of spectral weight observed for the $a$ axis
\textcolor{black}{with decreasing temperature}  (Fig.\,\ref{fig2} e) is a feature
of semiconductors, the \textcolor{black}{upturn seen in $\sigma_1(\omega)$ below
30 meV for the $c$ axis (Fig.\,\ref{fig2} f) signals metallicity}. Note,
however, that $\sigma_1(\omega)$ for the $a$ axis remains finite even at the
lowest temperatures and frequencies, i.e.\ there is no fully developed energy
gap. Thus we conclude that along the $a$ and $c$ axis CeRu$_4$Sn$_6$ behaves
predominantly insulator- and metal-like, \textcolor{black}{respectively}.

\begin{figure}[b!]
\centerline{\includegraphics[width=10cm,clip=true]{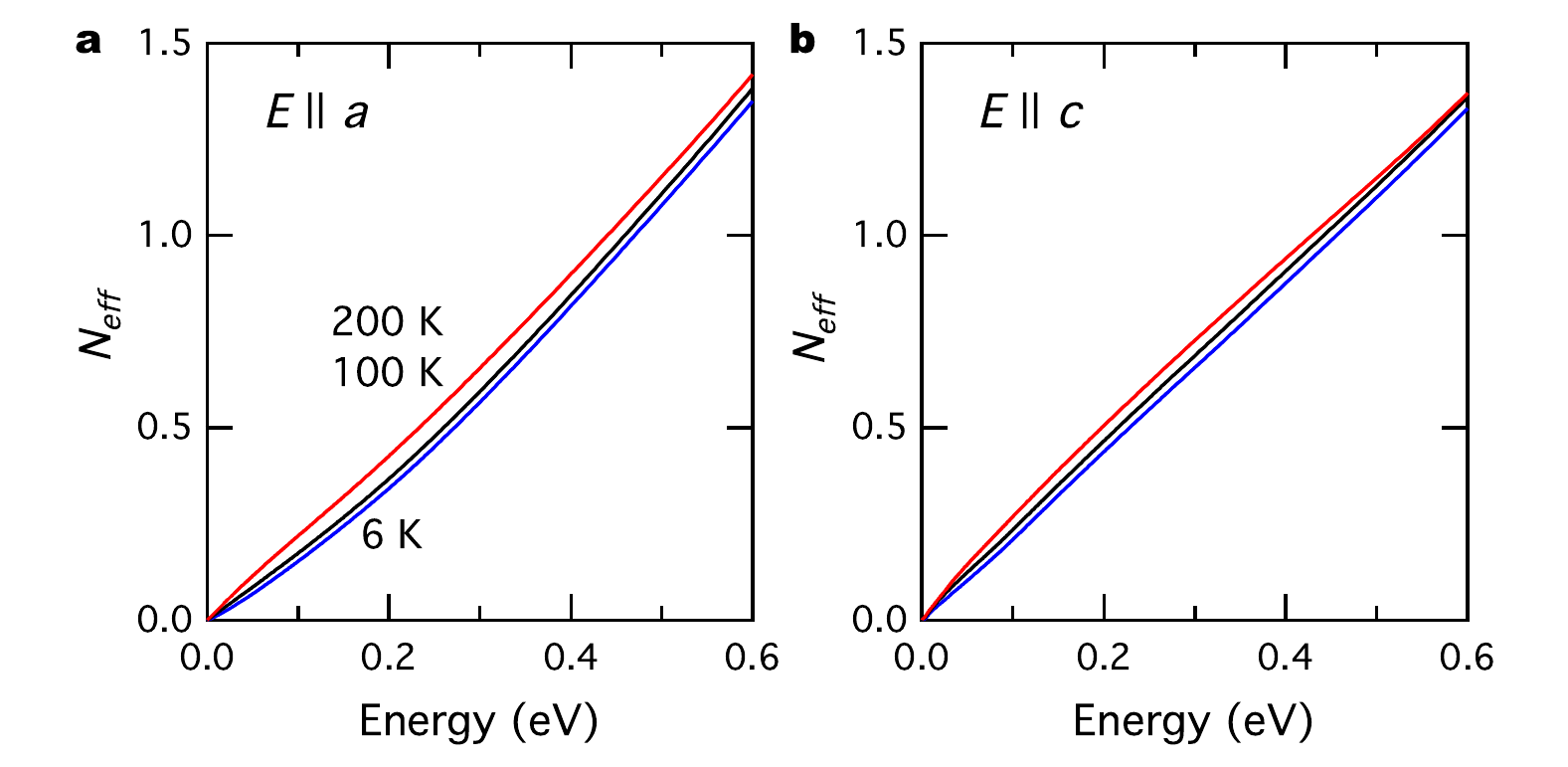}}
\vspace{-0.3cm}

\caption{(Color online) 
Integrated spectral weight of single-crystalline CeRu$_4$Sn$_6$ at 6, 100 and
200~K for $E \parallel a$ (a) and $E \parallel c$ (b).} 
\label{fig3} 
\end{figure}

The temperature evolution of the integrated spectral weight $N_{\rm
eff}(\omega)$ for the $a$ axis (Fig.\,\ref{fig3} a) suggests that the
\textcolor{black}{pseudo}gap is due to strong correlations as opposed to band
effects: The spectral weight lost at low temperatures \textcolor{black}{and
energies below 0.1 eV} due to the \textcolor{black}{pseudo}gap formation is still
not fully recovered at 0.6~eV, as similarly seen in other Kondo insulators
\cite{bucher94a,schlesinger93a}. Interestingly, this effect, though smaller, can
also be discerned for the $c$ axis (Fig.\,\ref{fig3} b). This indicates that
also there a remnant of a Kondo insulating gap exists.

\begin{figure}[b!]
\centerline{\includegraphics[width=\columnwidth,clip=true]{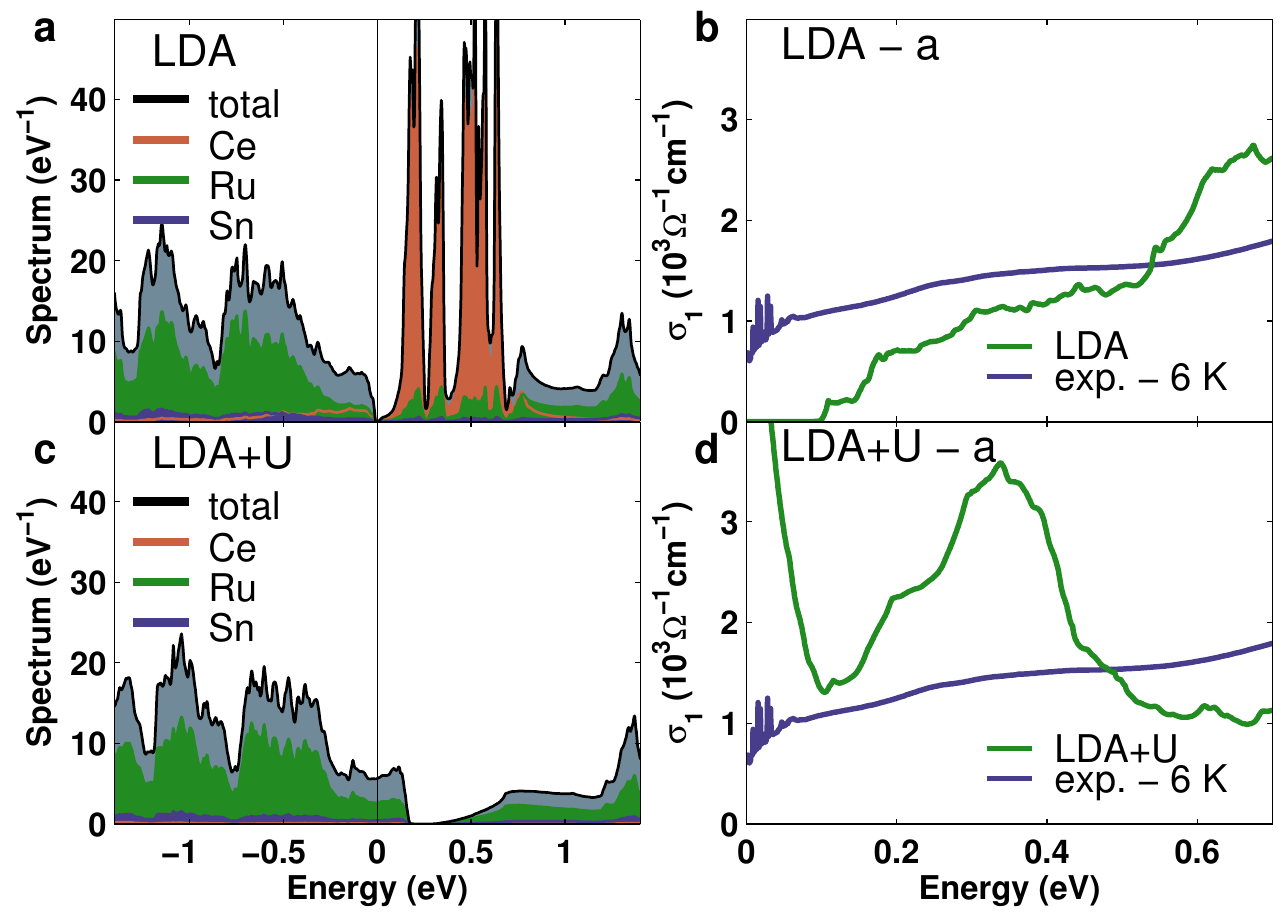}}
\vspace{-0.2cm}

\caption{(Color online) DOS and corresponding $a$-axis $\sigma_1(\omega)$
obtained by LDA (a, b) and LDA+U (c, d). The calculated $c$-axis conductivities
are similar (not shown).}
\label{fig4}
\end{figure}

\begin{figure*}
\centerline{\includegraphics[width=\textwidth]{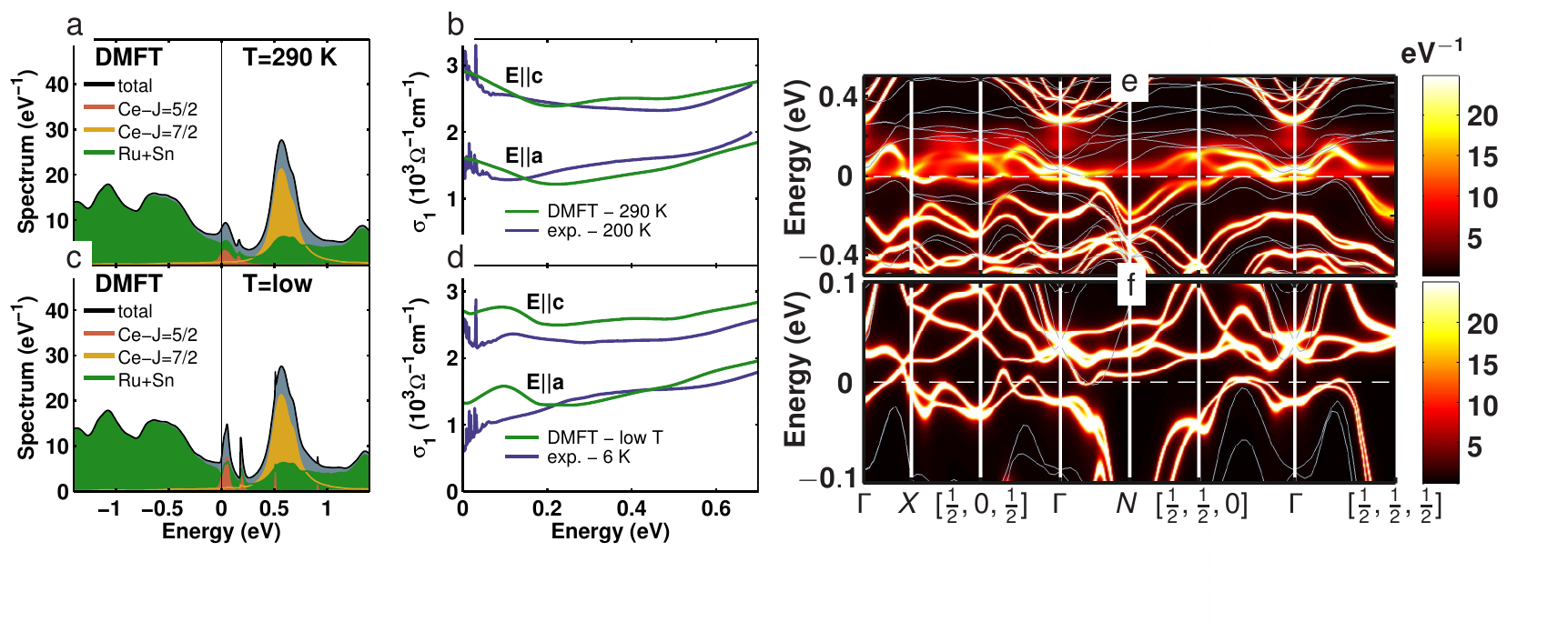}}
\vspace{-0.6cm}

\caption{(Color online) LDA+DMFT results: DOS at 290 K (a) and in the low
temperature limit (c), and the corresponding optical conductivities (b, d; c axis conductivities shifted by $+10^3 \Omega^{-1}$cm$^{-1}$ for clarity) and $k$-resolved spectra (e, f; LDA band structure displayed as thin lines), showing a band crossing at the Fermi level for $\Gamma\rightarrow X$ (c axis) and a gap in the $a - a$ plane.}
\label{fig5}
\end{figure*}

The only other non-cubic Kondo insulator for which $\sigma_1(\omega)$ results
along the different crystallographic directions are available is
CeFe$_2$Al$_{10}$ \cite{Kim11.2}. Here a Drude-like feature appears for all
three crystallographic directions; the anisotropy is thus much less pronounced
than for CeRu$_4$Sn$_6$.

Can the salient features of CeRu$_4$Sn$_6$ be theoretically understood and if so
on which level of approximation? To answer this question we performed, in a
first step, LDA band structure calculations (Fig.\,\ref{fig4} a). They yield a direct (but no indirect) bandgap of about 0.1~eV, separating Ru-$4d$ and Sn-$5p$
states below the Fermi level from Ce-$4f$ states above it. The corresponding
calculated $\sigma_1(\omega)$ is in strong disagreement with experiment
(Fig.\,\ref{fig4} b). In LDA+U, the $4f$-states are split by the Coulomb
repulsion $U$ so that one electron is transferred from Ru-$4d$ and Sn-$5p$ to
the Ce-$4f$ orbital below the LDA+U Fermi level. Hence, part of the Ru- and
Sn-states are now above the Fermi level (Fig.\,\ref{fig4} c) and CeRu$_4$Sn$_6$
is predicted to be metallic in all directions (Fig.\,\ref{fig4} d), again in
contrast to experiment.

In a third step, LDA + DMFT calculations were performed at different
temperatures. At high temperatures ($\sim 1000$~K) the spectrum is similar to
that of LDA+U (not shown). Upon reducing the temperature, we note the emergence of a Kondo resonance (red $f$-electron peak at the Fermi level in Fig.\,\ref{fig5} a, c). This has dramatic
consequences for $\sigma_1(\omega)$, which now shows good agreement with
experiment (Fig.\,\ref{fig5} b). Since the lowest experimental temperatures are
not accessible in our theoretical approach, we mimic them by switching off the
imaginary part of the self energy (Fig.\,\ref{fig5} c). This turns out to further enhance the anisotropy. In particular, $\sigma_1(\omega)$ decreases sizably towards the
lowest frequencies for the $a$ axis while it levels out for the $c$ axis
(Fig.\,\ref{fig5} d), thus correctly reproducing the experimental trends.

The $k$-resolved LDA+DMFT spectrum reveals the origin of the anisotropy. At 290 K, the Ce-4$f$ states are strongly broadened (smearing of bands around the Fermi level, Fig.\,\ref{fig5} e), which
indicates the vicinity of the Kondo temperature where the scattering rate is
high. At low temperatures, mimicked as above, the picture becomes clearer
(Fig.\,\ref{fig5} f): In large parts of the Brillouin zone there is a direct gap
reminiscent of a Kondo insulator, particularly within the tetragonal $a-a$
plane, see e.g.\ $\Gamma\rightarrow$\ ($\frac{1}{2}$,$\frac{1}{2}$,0). In contrast, in the
$c$ direction ($\Gamma\rightarrow X$) there is no such gap and the system is expected to
show characteristics of a heavy fermion metal.

In summary, we have investigated the optical properties of single-crystalline
CeRu$_4$Sn$_6$ by optical reflectivity measurements and LDA + DMFT calculations.
The experimentally observed anisotropy is very pronounced, with metal-like
features along one direction but semiconductor-like features elsewhere, and can be
traced back to the peculiar $k$ dependence of the correlated electron bands. The
weak metallicity of CeRu$_4$Sn$_6$ is thus clearly a bulk effect and not due to
topologically protected metallic surface states \cite{Dze10.1}. It will be most
enlightening to see whether magnetic Ising anisotropy goes along with this
peculiar quasiparticle anisotropy and whether any relation to ``hastatic order''
in URu$_2$Si$_2$ \cite{Cha12.1} can be established.

We thank M.\ Baenitz, M.\ Brando, A. T\'oth, and A.\ Pimenov for fruitful
discussions, A.\ Irizawa for sharing beam time at UVSOR, and K.\ Imura for
technical assistance. V.G.\ benefited from financial support from the Alexander
von Humboldt Foundation, P.W.\ and K.H.\ from the Austrian Science Fund (FWF-SFB
ViCoM F41), H.W., A.P.\ and S.P.\ from the FWF (project I623-N16) and the
European Research Council (ERC advanced grant 227378), and A.M.S\ from the DFG(OE 511/1-1) and the Science Faculty of UJ. Calculations were done on
the Vienna Scientific Cluster.

\bibliographystyle{prsty}

\begin{thebibliography}{10}
\bibitem{THz}
In the THz frequency range $R(\omega)$ is very close to unity where small uncertainties in $R$ can cause large errors in $\sigma_1$. Therefore, $\sigma_1(\omega)$ was obtained by a weighted average over finite frequency ranges.
\end{thebibliography}

\end{document}